%% file: 0-main.tex
\documentclass[10pt,conference]{IEEEtran}
\IEEEoverridecommandlockouts
\usepackage{cite}
\usepackage{amsmath,amssymb,amsfonts}
\usepackage{algorithmic}
\usepackage{graphicx}
\usepackage{textcomp}
\usepackage{xcolor}
\usepackage{booktabs}
\usepackage{xspace}
\usepackage{amssymb,amsmath}
\usepackage{flushend}

\newboolean{showcomments}
\setboolean{showcomments}{true}
\ifthenelse{\boolean{showcomments}}
 { \newcommand{\mynote}[2]{
      \fbox{\bfseries\sffamily\scriptsize#1}
        {\small$\blacktriangleright$\textsf{\emph{#2}}$\blacktriangleleft$}}}
        { \newcommand{\mynote}[2]{}}

\definecolor{DarkOrange}{rgb}{0.8,0.3,0.0}
\definecolor{DarkCyan}{rgb}{0.0, 0.55, 0.55}

\def\BibTeX{{\rm B\kern-.05em{\sc i\kern-.025em b}\kern-.08em
    T\kern-.1667em\lower.7ex\hbox{E}\kern-.125emX}}
\begin{document}

\title{{\sc iBiR}: Bug Report driven Fault Injection}

\newcommand{\toolname}{{\sc iBiR}\xspace}

\newcommand{\changing}[1]{{\color{black} #1}} 

\author{

\IEEEauthorblockN{Ahmed Khanfir, Anil Koyuncu, Mike Papadakis, Maxime Cordy, Tegawend\'e F. Bissyand\'e, \\Jacques Klein and Yves Le Traon}

\IEEEauthorblockA{
\textit{SnT, University of Luxembourg, Luxembourg}\\
\{firstname.surname\}@uni.lu
}
}


\maketitle

\begin{abstract}

Much research on software engineering and software testing relies on experimental studies based on fault injection. 
Fault injection, however, is not often relevant to emulate real-world software faults since it ``blindly'' injects large numbers of faults. It remains indeed challenging to inject few but realistic faults that target a particular functionality in a program. In this work, we introduce \toolname{}, a fault injection tool that addresses this challenge by exploring change patterns associated to user-reported faults. To inject realistic faults, we create mutants by re-targeting a bug report driven automated program repair system, i.e., reversing its code transformation templates. \toolname is further appealing in practice since it requires deep knowledge of neither of the code nor the tests, but just of the program's relevant bug reports. Thus, our  approach focuses the fault injection on the feature targeted by the bug report. We assess \toolname by considering the Defects4J dataset. 
Experimental results show that our approach outperforms the fault injection performed by traditional mutation testing in terms of semantic similarity with the original bug, when applied at either system or class levels of granularity, and provides better, statistically significant, estimations of test effectiveness (fault detection). Additionally, when injecting 100 faults, \toolname injects faults that couple with the real ones in \changing{36\%} of the cases, while mutants from mutation testing inject less than \changing{1\%}. Overall, \toolname targets real functionality and injects realistic and diverse faults.\\
\end{abstract}


\input{1-introduction}

\input{2-background}
\input{3-approach}
\input{4-rqs}
\input{5-setup}
\input{6-results}
\input{7-Threats}
\input{8-RW}
\input{9-conclusion}

\bibliographystyle{IEEEtran}
\bibliography{./bibliography/sample-base,./bibliography/refs-repair}


\end{document}

%% file: 1-introduction.tex
\section{Introduction}
\label{sec:introduction}
A key challenge of fault injection techniques (such as mutation analysis) is to emulate the effects of real faults. This property of representativeness of the injected faults is of particular importance since fault injection techniques are widely used by researchers when evaluating and comparing bug finding, testing and debugging techniques, e.g., test generation, bug fixing, fault localisation, etc, \cite{PapadakisK00TH19}. This means that there is a high risk of mistakenly asserting test effectiveness in case the injected faults are non-representative.
 
Typically, fault injection techniques introduce faults by making syntactic changes in the target programs’ code using a set of simple syntactic transformations \cite{DeMilloLS78, NatellaCDM13, LanzaroNWCS15}, usually called mutation operators. These transformations have been defined based on the language syntax \cite{0020331} and are ``blindly'' mutating the entire codebase of the projects, injecting large numbers of mutants, with the hope to inject some realistic faults. This means that there is a limited control on the fault types and the locations where to inject faults. In other words, the appropriate ``what’’ and ``where'' to inject faults in order to make representative fault injection has been largely ignored by existing research. 

Fault injection techniques may also draw on recent research that mines fault patterns \cite{TufanoWBPWP19, BrownVLR17} and demonstrate some form of realism w.r.t. real faults. These results are encouraging because they indicate that the injected faults may carry over the realism of the patterns. This may remove a potential validity threat, but at the same time, it is limited as it does not provide any control on the locations and target functionality, thus impacting fault representativeness \cite{NatellaCDM13, ChekamPBTS20, PapadakisSYB18}. 

This is an important limitation especially for large real-world systems because of the following two reasons: a) injecting faults everywhere escalates the application cost due to the large number of mutants introduced
 and b) the results could be misleading since a tiny ratio of the injected faults are coupled to the real ones \cite{PapadakisSYB18} and the injected set of faults do not represent the likelihood of faults appearing in the field \cite{NatellaCDM13}.  
Therefore, representativeness of the injected faults in terms of fault types and locations is of outmost importance w.r.t. both application cost and accuracy of the method.


To bypass these issues, one could use real faults (mined from the projects’ repositories) or directly apply the testing approach to a set of programs and manually identify potential faults. While such a solution brings realism into the evaluations, it is often limited to few fault instances (of limited diversity), requires expensive manual effort in identifying the faults and fails to offer the experimental control required by many evaluation scenarios. 

We advance in this research direction by bringing realism in the fault injection via leveraging information from bug reports. Bug reports often include sufficient information for debugging techniques in order to localize \cite{zhou2012should}, debug \cite{papadakis2015metallaxis} and repair faults \cite{koyuncu2019ifixr} that happened in the field. Therefore, together with specially crafted defect patterns (mined through systematic examination of real faults) such information can guide fault injection to target critical functionality, mimic real faulty behaviour and make realistic fault injection. Perhaps more importantly, the use of bug reports removes the need for knowledge of the targeted system or code.



Our method starts from the target project and a bug report written in natural language. It then applies Information Retrieval (IR)-based fault localization \cite{zhou2012should} in order to identify the relevant places where to inject faults. It then injects recurrent fault instances (fault patterns) that were manually crafted using a systematic analysis of frequent bug fixes, prioritized according to their position and type. This way our method performs fault injection, using realistic fault patterns, by targeting the features described by the bug reports. 

We implemented our approach in a system called \toolname and evaluated its ability to imitate \changing{157} real faults. In particular we evaluated a) the semantic similarity of real and injected faults, b) the coupling\footnote{Injected faults couple with the real ones when injected faults are detected only by test cases that detect the real faults. This implies that the injected faults provide good indications on whether tests are capable of detecting the coupled faults.} relation between injected and real faults, and c) the ability of the injected faults to indicate test effectiveness (fault detection) 
when tested with different test suites. Our results show that \toolname manages to imitate the targeted faults, with a median semantic similarity value of \changing{0.58}, which is significantly higher than the \changing{0.0} achieved by using traditional mutation testing, when injecting the same number of faults. 

Interestingly, we found that \toolname injects faults that couple with the real ones in \changing{36\%} of the targeted cases. This is achieved by injecting 100 faults per target (real) fault and it is approximately \changing{50} times higher than the coupled mutants produced by mutation testing. Fault coupling is one of the most important testing properties \cite{PapadakisCT18, KintisPPVMT18}, here indicating that one can use the injected faults instead of the real ones. 

Another key finding of our study is that the injected faults provide much better indication on test effectiveness (fault detection) than mutation testing as their detection ratios discriminate between actual failing and passing test suites, while mutant detection rates cannot.  
This implies that the use of \toolname yields more accurate results than the use of traditional mutation testing. 

Overall, our primary contributions are: 

\begin{itemize}
    \item We introduce the notion of bug report-driven fault injection. Bug reports can be used to inject realistic faults. 
    
    \item We introduce a set of mutation operators based on frequently used patch patterns that are reverted to inject realistic faults. 
    
 \item We present \toolname, an automatic fault injection method, which is driven by bug reports to emulate real faults.
  
  \item We provide empirical evidence demonstrating that \toolname outperforms the current state of practice in mutation testing w.r.t. fault representativeness and coupling. 
  
\end{itemize}

%% file: 2-background.tex
\section{Background}
\label{sec:background}

\subsection{Fault Localization}

Fault localization is the activity of identifying the 
suspected fault locations, which will be 
transformed to generate patches. Several automated fault localization techniques have been proposed~\cite{wong2016survey}, such as slice-based~\cite{wong2010family}, spectrum-based~\cite{abreu2009spectrum}, statistics-based~\cite{liblit2005scalable}, mutation-based \cite{papadakis2015metallaxis} and etc. 

Fault localisation techniques based on Information Retrieval (IR)~\cite{deerwester_indexing_1990,frakes_information_1992,manning_foundations_1999,salton_introduction_1986} 
exploit textual bug reports to identify code chunks relevant to the bug, without relying on test cases. IR-based fault localisation tools extract tokens from the bug report to formulate a query to be matched with the collection of documents formed by the source code files
~\cite{lukins2010bug,saha2013improving,wang2014version,wen2016locus,youm2015bug,zhou2012should}. Then, they rank the documents based on their relevance to the query, such that source files ranked higher are more likely to contain the fault. Recently, automated program repair methods have been designed on top of IR-based fault localization \cite{koyuncu2019ifixr}. They achieve comparable performance to methods using spectrum-based fault localization, yet without relying on the assumption that test cases are available.

We leverage IR-based fault localization to achieve a different goal: instead of localising the reported bug, we aim at \emph{injecting faults} at code locations that implement a functionality similar to the fault targeted by the bug report description. 

\subsection{Mutation Testing}

Mutation testing is a popular fault-based testing technique \cite{PapadakisK00TH19}. It operates by inserting artificial faults into a program under test, thereby creating many different versions (named \emph{mutants}) of the program. The artificial faults are injected through syntactic changes to all program locations in the original program, based on predefined rules named \emph{mutation operators}. Such operators can, for instance, invert relational operators (e.g., replacing $\geq$ with $<$). 

Mutants can be used to indicate the strengths of test suites, based on their ability to distinguish the mutants from the original program. If there exists a test case distinguishing the original program from a particular mutant, then the mutant is said to be \textit{killed}. Then, we term a mutant to be ``coupled'' with respect to a particular fault if the test cases that kill it are a subset of the test cases that can also detect that fault (make the program fail by exerting the fault).

Previous research has shown that the choice of mutation operators and location can affect the fault-revealing ability of the produced mutants \cite{Andrews:TSE:2006, LPKHTV17}. Thus, it is important to select appropriate mutation testing strategies. Nevertheless, previous research has shown that random mutant sampling achieves comparable results with the mutation testing state of the art \cite{KurtzAODKG16, ChekamPBTS20}, making the random mutant sampling a natural baseline to compare with. 

Another issue involved in mutation testing campaigns is the application cost of the method. The problem stems from the vast number of faults that are injected, which need to be executed with a large number of test suites, thereby escalating the computational demands of the method \cite{PapadakisK00TH19}. Unfortunately, the mutant execution problem becomes intractable when test execution is expensive or the test suites involve system level tests, thereby often limiting mutation testing application to unit level. This is a major problem when performing fault tolerance \cite{NatellaCDM13}, or other large-scale testing campaigns. Luckily, recent studies have shown that only a tiny number of the injected faults are useful \cite{PapadakisSYB18, ammann_establishing_2014, KurtzAODKG16}, suggesting that a handful number of injected faults should be sufficient to perform testing. Though, it remains an open question on how to identify them. 

We fill this gap, by using bug report-driven fault injection. In essence we leverage IR-based fault localization techniques to identify the locations where fault injection should happen, i.e., locations relevant to the targeted functionality described in the bug report, and apply frequent fault patterns to produce mutants that behave similar to real faults.

\subsection{Fix Patterns}

In automated program repair \cite{GouesPR19}, a common way to generate patches is to apply fix patterns~\cite{kim2013automatic} (also named fix templates~\cite{liu2018mining} or program transformation schemes~\cite{hua2018towards}) in suspicious program locations (detected by fault localization). Patterns used in the literature~\cite{kim2013automatic,saha2017elixir,durieux2017dynamic,
liu2018mining,hua2018towards,koyuncu2018fixminer,martinez2018ultra,liu2019avatar,liu2018mining2} have been defined manually or automatically (mined from bug fix datasets). 

Instead of fix patterns, we use \emph{fault patterns} that are fix patterns inverted. Since fix patterns were designed using recurrent faults their related fault patterns introduce them. This helps injecting faults that are similar to those described in the bug reports. \toolname inverts and uses the patterns implemented by \emph{TBar}~\cite{liu2019tbar} as we detail in the following Section. 

\begin{figure*}[t]
\centering 
    \vspace{-1.0em}
    \includegraphics[width=\textwidth]{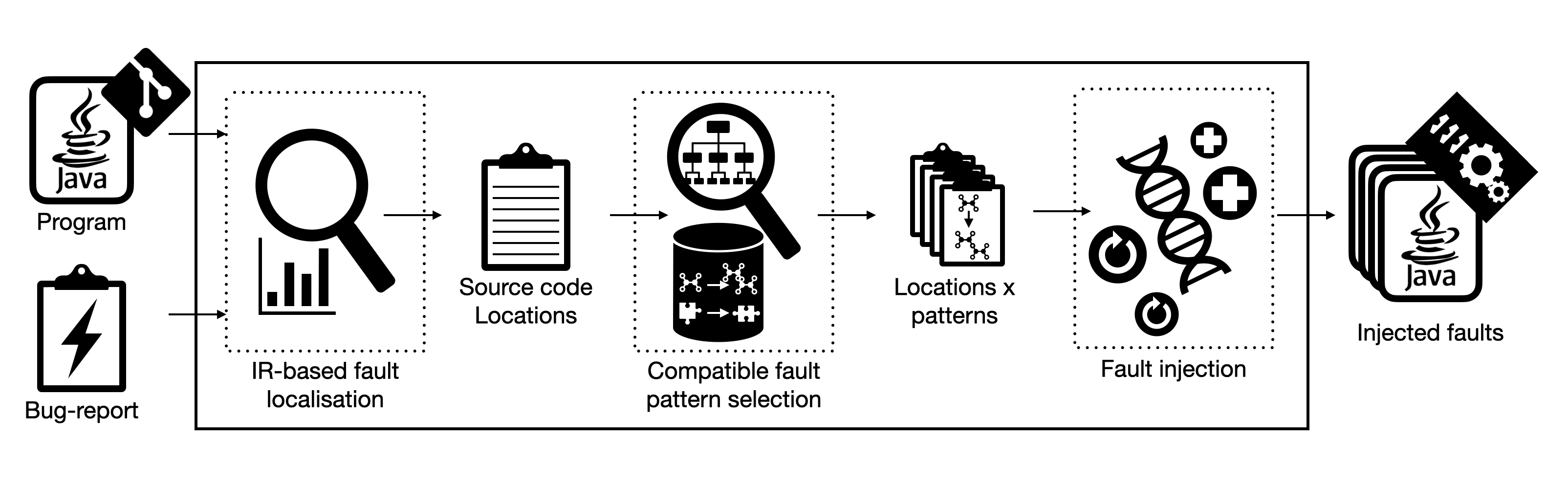}
     \vspace{-3.5em}
    \caption{The \toolname fault injection workflow.}
    \label{fig:pipeline}
    \vspace{-0.8em}
\end{figure*}


%% file: 3-approach.tex
\section{Approach}
\label{sec:approach}

We propose \toolname, the first fault injection approach that utilizes information extracted from bug reports to emulate real faults. A high level view of the way \toolname works is shown in Figure \ref{fig:pipeline}. Our approach takes as input (1) the source code of the program of interest and (2) a resolved bug report of that program, written in natural language. The objective is to inject artificial faults in the program (one by one, creating multiple faulty versions of the program) that imitate the original bug. To do so, \toolname proceeds in three steps.

First step:  \toolname identifies relevant locations to inject the faults. It applies IR-based fault localization to determine, from the bug report, the code locations (statements) that are likely to be relevant to the target fault. These locations are ranked according to their likelihood to be the feature described by the bug report, hence are relevant to inject faults.

Second step: \toolname applies fault patterns on the identified code locations. We build our patterns by inverting fix patterns used in automated program repair approaches~\cite{liu2019tbar}. Our intuition is that, since fix patterns are used to fix bugs, inverted patterns may introduce a fault similar to the original bug. For each location, we apply only patterns that are syntactically compatible with the code location. This step yields a set of faults to inject, i.e., pairs composed of a location and a pattern.

Third step:  our method ranks the location-pattern pairs wrt. the location likelihood and priority order of the patterns. 
Then \toolname takes each pair (in order) and applies the pattern to the location, injecting a fault in the program. 
We repeat the process until the desired number of injected faults has been produced or until all location-pattern pairs have been considered.





\subsection{Bug Report driven Fault Localization}

IR-based fault localization (IRFL)~\cite{parnin2011automated,wang2015evaluating} leverages potential similarity between the terms used in a bug report and the program source code to identify relevant buggy code locations. It typically starts by extracting tokens from a given bug report to formulate a {\em query} to be matched in a search space of {\em documents} formed by the collections of source code files and indexed through tokens extracted from source code~\cite{zhou2012should,wen2016locus,wong2014boosting,saha2013improving,wang2014version,lukins2010bug}. IRFL approaches then rank the documents based on a probability of relevance. 
Top-ranked files are likely to contain the buggy code.

We follow the same principle to identify promising locations where to inject realistic faults. We rely on the information contained in the bug report to localize the code location with the highest similarity score. Most IRFL techniques have focused on file-level localization, which is too coarse-grained for our purpose of injecting fault. Thus, we rather use a statement-level IRFL approach that has been successfully applied to support program repair~\cite{koyuncu2019ifixr}.

It is to be noted that, contrary to program repair, we do not aim to identify the exact bug location. We are rather interested in locations that allow injecting realistic faults (similar to the bug). This means that IRFL may pinpoint multiple locations of interest for fault injection even if those were not buggy code locations.

\subsection{Fault patterns}

\input{patterns-taxonomy-table}



We start from the fix patterns developed in TBar \cite{liu2019tbar}, a state of the art pattern-based program repair tool. Any pattern is described by a context, i.e., an AST node type to which the pattern applies, and a recipe, a syntactical modification to be performed. For each pattern, we define a related fault injection pattern that represents the inverse of that pattern. For instance, inverting the fix pattern that consists of adding an arbitrary statement yields a \emph{remove statement} fault pattern. Interestingly, some fix patterns are symmetric in the sense that their inverse pattern is also a fix pattern, e.g., inverting a Boolean connector. These patterns can thus be used for both bug fixing and fault injection. Table~\ref{tab:patterns} enumerates the resulting set of fault injection patterns used by our approach.



Given a location (code statement) to inject a fault into, we identify the patterns that can be applied to the statement. To do so, our method starts from the AST node of the statement and visits it exhaustively, in a breadth-first manner. Each time it meets an AST node that matches the context of a fault pattern, it memorizes the node and the pattern for later application. Then the method continues until it has visited all AST nodes under the statement node. This way, we enumerate all possible applications of all fault patterns onto the location.

Since more than one pattern may apply to a given location, we prioritize them by leveraging heuristic priority rules previously defined in automated program repair methods (these were inferred from real-world bug occurrences \cite{liu2019tbar}). This means that every fault injection pattern gets the priority order of its inverse fix pattern.


\subsection{Fault injection}

The last step consists of applying, one by one, the fault patterns to inject faults at the program locations identified by IRFL. Locations of higher ranking are considered first. Within a location, pattern applications are ordered based on the pattern priority. By applying a pattern to a corresponding AST node of the location, we inject a fault within the program before recompiling it. If the program does not compile, we discard the fault and restart with the next one. We continue the process until it reaches the desired number of (compilable) injected faults or all locations and patterns have been considered.

%% file: patterns-taxonomy-table.tex
\begin{table*}
\vspace{-1.0em}
\caption{iBIr fault injection patterns.}
\vspace{-0.5em}
\label{tab:patterns}
\begin{tabular}{@{}llll@{}}
\toprule
\textbf{Pattern context category} &
  \textbf{Bug injection pattern} &
  \textbf{example input} &
  \textbf{example output} \\ \midrule
\textbf{Insert Statement} &
  \begin{tabular}[c]{@{}l@{}}Insert a method call, \\ before or after the localised statement.\end{tabular} &
  \textit{someMethod(expression);} &
  \textit{\begin{tabular}[c]{@{}l@{}}someMethod(expression);\\ method(expression);\end{tabular}} \\
\textbf{} &
  \begin{tabular}[c]{@{}l@{}}Insert a return statement, \\ before or after the localised statement.\end{tabular} &
  \textit{statement;} &
  \textbf{\begin{tabular}[c]{@{}l@{}}statement;\\ return VALUE;\end{tabular}} \\
\textbf{} &
  Wrap a statement with a try-catch. &
  \textit{statement;} &
  \textit{\begin{tabular}[c]{@{}l@{}}try\{\\       statement;\\ \} catch (Exception e)\{ ... \}\end{tabular}} \\
\textbf{} &
  \begin{tabular}[c]{@{}l@{}}Insert an if checker: wrap a \\ statement with an if block.\end{tabular} &
  \textit{statement;} &
  \textit{\begin{tabular}[c]{@{}l@{}}if (conditional\_exp) \{\\       statement; \}\end{tabular}} \\ \midrule
\textbf{Mutate Class Instance Creation} &
  \begin{tabular}[c]{@{}l@{}}Replace an instance creation call by \\ a cast of the super.clone() method call.\end{tabular} &
  \textit{... new T();} &
  \textit{... (T) super.clone();} \\ \midrule
\textbf{Mutate Conditional Expression} &
  Remove a conditional expression. &
  \textit{condExp1 \&\& condExp2} &
  \textit{condExp1} \\
\textbf{} &
  Insert a conditional expression. &
  \textit{condExp1} &
  \textit{condExp1 \&\& condExp2} \\
\textbf{} &
  Change the conditional operator. &
  \textit{condExp1 \&\& condExp2} &
  \textit{condExp1 \(\vert \vert\) condExp2} \\ \midrule
\textbf{Mutate Data Type} &
  Change the declaration type of a variable. &
  \textit{T1 var ...;} &
  \textit{T2 var ...;} \\
\textbf{} &
  Change the casting type of an expression. &
  \textit{... (T1) expression ...;} &
  \textit{... (T2) expression ...;} \\ \midrule
\textbf{Mutate float or double Division} &
  Remove a float or a double cast &
  \textit{... dividend / (float) divisor ...;} &
  ... dividend / divisor ...; \\
\textbf{} &
  from the divisor. &
  \textit{... intVarExp / 10d ...;} &
  \textit{... intVarExp / 10 ...;} \\
\textbf{} &
  Remove a float or a double cast &
  \textit{... (float) dividend / divisor ...;} &
  \textit{... dividend / divisor ...;} \\
\textbf{} &
  from the dividend. &
  \textit{... 1.0 / var ...;} &
  \textit{... 1 / var ...;} \\
\textbf{} &
  Replace float or double multiplication &
  \textit{... (1.0 / divisor) * dividend ...} &
  \textit{... dividend / divisor ...;} \\
\textbf{} &
  by an int division. &
  \textit{... 0.5 * intVarExp ...;} &
  \textit{... intVarExp / 2 ...;} \\ \midrule
\textbf{Mutate Literal Expression} &
  \begin{tabular}[c]{@{}l@{}}Change boolean, number or string \\ literals in a statement by another literal\\ or expression of the same type.\end{tabular} &
  \textit{\begin{tabular}[c]{@{}l@{}}... string\_literal1 ... \\ \\ ... int\_literal ...\end{tabular}} &
  \textit{\begin{tabular}[c]{@{}l@{}}... string\_literal2 ...\\ \\ ... int\_expression ...\end{tabular}} \\ \midrule
\textbf{Mutate Method Invocation} &
  Replace a method call by another one. &
  \textit{... method1(args) ...} &
  \textit{... method(args) ...} \\
\textbf{} &
  Replace a method call argument by another one. &
  \textit{... method(arg1, arg2) ...} &
  \textit{... method(arg1, arg3) ...} \\
\textbf{} &
  Remove a method call argument. &
  \textit{... method(arg1, arg2) ...} &
  \textit{... method(arg1) ...} \\
\textbf{} &
  Add an argument to a method call &
  \textit{... method(arg1) ...} &
  \textit{... method(arg1, arg2) ...} \\ \midrule
\textbf{Mutate Return Statement} &
  Replace a return experession by an other one. &
  \textit{return expr1;} &
  \textit{return exp2;} \\ \midrule
\textbf{Mutate Variable} &
  \begin{tabular}[c]{@{}l@{}}Replace a variable by another variable \\ or an expression of the same type.\end{tabular} &
  \textit{\begin{tabular}[c]{@{}l@{}}... var1 ...\\ ... var1 ...\end{tabular}} &
  \textit{\begin{tabular}[c]{@{}l@{}}... var2 ...\\ ... exp ...\end{tabular}} \\ \midrule
\textbf{Move Statement} &
  Move a statement to another position. &
  \textit{\begin{tabular}[c]{@{}l@{}}statement;\\ ...\end{tabular}} &
  \textit{\begin{tabular}[c]{@{}l@{}}...\\ statement;\end{tabular}} \\ \midrule
\textbf{Remove Statement} &
  Remove a statement. &
  \textit{\begin{tabular}[c]{@{}l@{}}statement;\\ ...\end{tabular}} &
  \textit{...} \\
\textbf{} &
  Remove a method. &
  \textit{method(args)\{ statement; \}} &
  \textit{...} \\ \midrule
\textbf{Mutate Operators} &
  Replace an Arithmetic operator. &
  \textit{... a + b ...} &
  \textit{... a - b ...} \\
\textbf{} &
  Replace an Assignment operator. &
  \textit{... c += b ...} &
  \textit{... c -= b ...} \\
\textbf{} &
  Replace a Relational operator. &
  \textit{... a \textless b ...} &
  \textit{... a \textgreater b ...} \\
\textbf{} &
  Replace a Conditional operator. &
  \textit{... a \&\& b ...} &
  \textit{... a \(\vert \vert\) b ...} \\
\textbf{} &
  Replace a Bitwise or a Bit Shift operator. &
  \textit{... a \& b ...} &
  \textit{... a \(\vert\) b ...} \\
\textbf{} &
  Replace an Unary operator. &
  \textit{a++} &
  \textit{a\text{-}\text{-}} \\
\textbf{} &
  Change arethmetic operations order. &
  \textit{a + b * c} &
  \textit{c + b * a} \\ \bottomrule
\end{tabular}
\vspace{-1.0em}
\end{table*}

%% file: 4-rqs.tex
\section{Research Questions}
\label{sec:rqs}

Our approach aims at injecting faults that imitate real ones by leveraging the information included in bug reports. Therefore, a natural question to ask is how well \toolname's faults imitate the targeted (real) ones. Thus, we ask: 

\begin{description}
\item[RQ1] \emph{(Imitating bugs):} Are the \toolname faults capable of emulating, in terms of semantic similarity, the targeted (real) ones? 
\end{description}

To answer this question, we check whether any of the injected faults imitate well the targeted ones. Following the recommendations from the mutation testing literature \cite{PapadakisSYB18} we approximate the program behaviour through the project test suites and compare the behaviour similarity of the test cases w.r.t. their pass and failing status using the Ochiai similarity coefficient. This is a typical way of computing the semantic similarity of mutants and faults in mutation-based fault localization \cite{papadakis2015metallaxis, MoonKKY14}. 

We then turn our attention to the similarity of the injected fault sets and contrast them with mutants such as those used by modern mutation testing tools \cite{KintisPPVMT18}. Hence we ask:

\begin{description}
\item[RQ2] \emph{(Comparison with mutation testing):} How does \toolname compare with mutation testing, in terms of semantic similarity? 
\end{description}

We answer this question by injecting mutants using the standard operators employed by mutation testing tools \cite{KintisPPVMT18} and measuring their semantic similarity with the targeted faults. To make a fair comparison, we inject the same number of faults per target. For \toolname we selected the top-ranked mutants while for mutation testing we randomly sampled mutants across the entire project codebase. Random mutant sampling forms our baseline since it performs comparably to the alternative mutant selection methods \cite{KurtzAODKG16, ChekamPBTS20}. Also, since we are interested in the relative differences between the injected fault sets, we repeat our experiments multiple times using the same number of faults (mutants). 

Our approach identifies the locations where bugs should be injected through an IR-based fault localization method. This may give significant advantages when applied at the project level, but these may not carry on individual classes. Such class level granularity level may be well suited for some test evaluation tasks, such as automatic test generation \cite{FraserA13}. To account for this, we performed mutation testing (using the traditional mutation operators) at the targeted classes (classes where the faults were fixed). To make a fair comparison we also restricted \toolname to the same classes and compared the same number of mutants. This leads us to the following question:

\begin{description}
\item[RQ3] \emph{(Comparison at the target class):} How does \toolname compare with mutation testing, in terms of semantic similarity, when restricted to particular classes?  
\end{description}

We answer this question by injecting faults in only the target classes using the \toolname bug patterns and the traditional mutation operators. Then we compare the two approaches the same way as we did in RQ1 and RQ2. 

Up  to this point, the answers to the posed questions provide evidence that using our approach yields mutants that are semantically similar to the targeted bugs. Although, this is important and demonstrates the potential of our approach, it does not necessarily mean that the injected faults are strongly coupled with the real ones\footnote{Mutants are coupled with real faults if they are killed only by test cases that also reveal the real faults}. Mutant and fault coupling is an important property for mutants that significantly helps testing \cite{JustJIEHF14}. Therefore, we seek to investigate:

\begin{description}
\item[RQ4] \emph{(Mutant and fault coupling):} How does \toolname compare with mutation testing with respect to mutant and fault coupling?
\end{description}

To answer this question we check whether the faults that we inject are detected only by the failing tests, i.e., only by the tests that also reveal the target fault. Compared to similarity metrics, this coupling relation is stricter and stronger. 

After answering the above questions we turn our attention to the actual use of mutants in test effectiveness evaluations. Therefore, we are interested in checking the correlations between the failure rates of the sets of the injected faults we introduce and the real ones. To this end, we ask:

\begin{description}
\item[RQ5] \emph{(Failure estimates):} Are the injected faults leading to failure estimates that are representative of the real ones? How do these estimates compare with mutation testing?
\end{description}

The difference of RQ5 from the other RQs is that in RQ5, a set of injected faults is evaluated while, in the previous RQs only isolated mutant instances.    

%% file: 5-setup.tex
\section{Experimental Setup}
\label{sec:setup}

\subsection{Dataset \& Benchmark}

To evaluate \toolname we needed a set of benchmark programs, faults and bug reports. We decided to use Defects4J~\cite{just2014defects4j} since it is a benchmark that includes real-world bugs and it is quite popular in software engineering literature. 

\subsubsection{Linking the bugs with their related reports} To identify which bug report describes a given bug
in the Defects4J, we followed the same process as in the study of Koyuncu et al.~\cite{koyuncu2019ifixr}. Unfortunately, it was not possible to link the bug reports with the defects for the Joda-Time, JFreeChart and Closure because their repositories and issue tracking systems have been migrated into GitHub without any mapping of the bug report identifiers. This means that in these projects the bug identifiers that were used in the commit are meaningless. We therefore decided to ignore these projects in an attempt to make our evaluation data as clean as possible. 

For the Lang and Math projects, we used the bug linking strategies that are implemented in the Jira issue tracking software and used the approach of Fischer et al. \cite{FischerPG03} and Thomas et al. \cite{ThomasNBH13} to map the sought bugs with the corresponding reports. Precisely, we crawled the relevant bug reports and checked their links.  We selected bug reports that were tagged as ``BUG'' and marked as 
``RESOLVED'' or ``FIXED'' and have a ``CLOSED'' status. Then we searched the commit logs to identify related identifiers (IDs) that link the commits with the corresponding bug. 

Our resulting bug dataset included the 171 faults of Defect4J related to the Lang and Math projects. We discarded \changing{10} defects because they had a bug report with undesired status in the bug tracking system, or there were issues with the buggy program versions such as missing files from the repository at the reporting time. We also discarded another \changing{4} defects because \toolname generated less than 5 mutants in total. This leaves us with 157 faults.

\subsection{Experimental Procedure}
To compare the fault injection techniques we need to set a common basis for comparison. We set this basis as the number of injected faults since it forms a standard cost metric \cite{OffuttLRUZ96} that puts the studied methods under the same cost level. We used sets of 5, 10, 30, and 100 injected faults since our aim is to equip researchers with few representative faults, per targeted fault, in order to reach reasonable execution demands. 

To measure how well the injected faults imitate the real ones (answer RQ1, RQ2 and RQ3) we use a semantic similarity metric (Ochiai coefficient) between the test failures on the injected and real (targeted) faults. This coefficient quantifies the similarity level of the program behaviours exercised by the test suites and is often used in mutation testing literature \cite{PapadakisSYB18}. The metric takes values in the range [0, 1] with 0 indicating complete difference and 1 exact match. We treated the injected faults that were not detected by any of the test suites as equivalent mutants \cite{AndrewsBLN06, PapadakisDT14}. This choice does not affect our results since we approximate the program behaviours through the projects test suites, i.e., they are never killed. 

To measure whether the injected faults couple with the existing ones (answer RQ4), we followed the process suggested by Just et al. \cite{JustJIEHF14} and identified whether there were any injected faults that were killed by at least one failing test (test that detects the real fault) and not by any passing test (test that does not detect the real fault). In RQ5 we randomly sampled 50 test suites, subsets of the accompanied test suites, that included between \changing{10\%} to \changing{30\%} test cases of the original test suite and recorded the ratios of the injected faults that are detected when injecting 5, 10, 30 and 100 faults. We also recorded binary variables indicating whether or not each test suite detects the targeted fault. This process simulates cases where test suites of different strengths are compared. Based on these data, we computed two statistical correlation coefficients, the Kendall and Pearson.  

To further validate whether the two approaches provide sufficient indicators on the effectiveness of the test suites, we check whether the detection ratios of the injected faults are statistically higher when test suites detect the targeted faults than when they do not. 

To reduce the influence of stochastic effects we used the Wilcoxon test with a significance level of 0.05. This helped deciding whether the differences we observe can be characterised as statistically significant. Statistical significance does not imply sizable differences and thus, we also used the Vargha Delaney effect size $\hat{\text{A}}_{12}$ \cite{Vargha00}. In essence, the $\hat{\text{A}}_{12}$ values quantify the level of the differences. For instance, a value $\hat{\text{A}}_{12} = 0.5$ can be interpreted as a tendency of equal value of the two samples. $\hat{\text{A}}_{12} > 0.5$ suggest that the first set has higher values, while $\hat{\text{A}}_{12} < 0.5$ suggest the opposite. 

\subsection{Implementation}
To perform our experiments we set the following parameters in our framework:
First, we limit the IR fault localization on the 20 top ranked suspicious files, per bug report. We then searched them for the exact statements where to inject faults. We also ensured that the IR engine is not trained with bug reports that we aim to localize. Second, for the mutation testing, denoted as ``Mutation'' in our experiments, we used randomly sampled mutants from those produced by typical mutation operators, coming from mutation testing literature. In particular we implemented the muJava intra-method mutation operators \cite{MaOK05}, which are the most frequently used \cite{KintisPPVMT18}. 
Third to reduce the noise from stillborn mutants, i.e., mutants that do not compile, we discarded without taking into any consideration, i.e., prior to our experiment, every mutant that did not compile or its execution with the test suite exceeded a timeout of \changing{5 minutes}. Fourth, when answering the RQ3, we found out that there were many cases where \toolname injected less than 100 faults. To perform a fair comparison, we discarded these cases (for both approaches). This means that we always report results where both studied approaches manage to inject the same number of faults. 

%% file: 6-results.tex
\section{Results}
\label{sec:results}

\subsection{RQ1: Semantic similarity between injected and real faults}

\begin{figure}[t]
\vspace{-1.0em}
\centering 
    \includegraphics[width=0.5\textwidth]{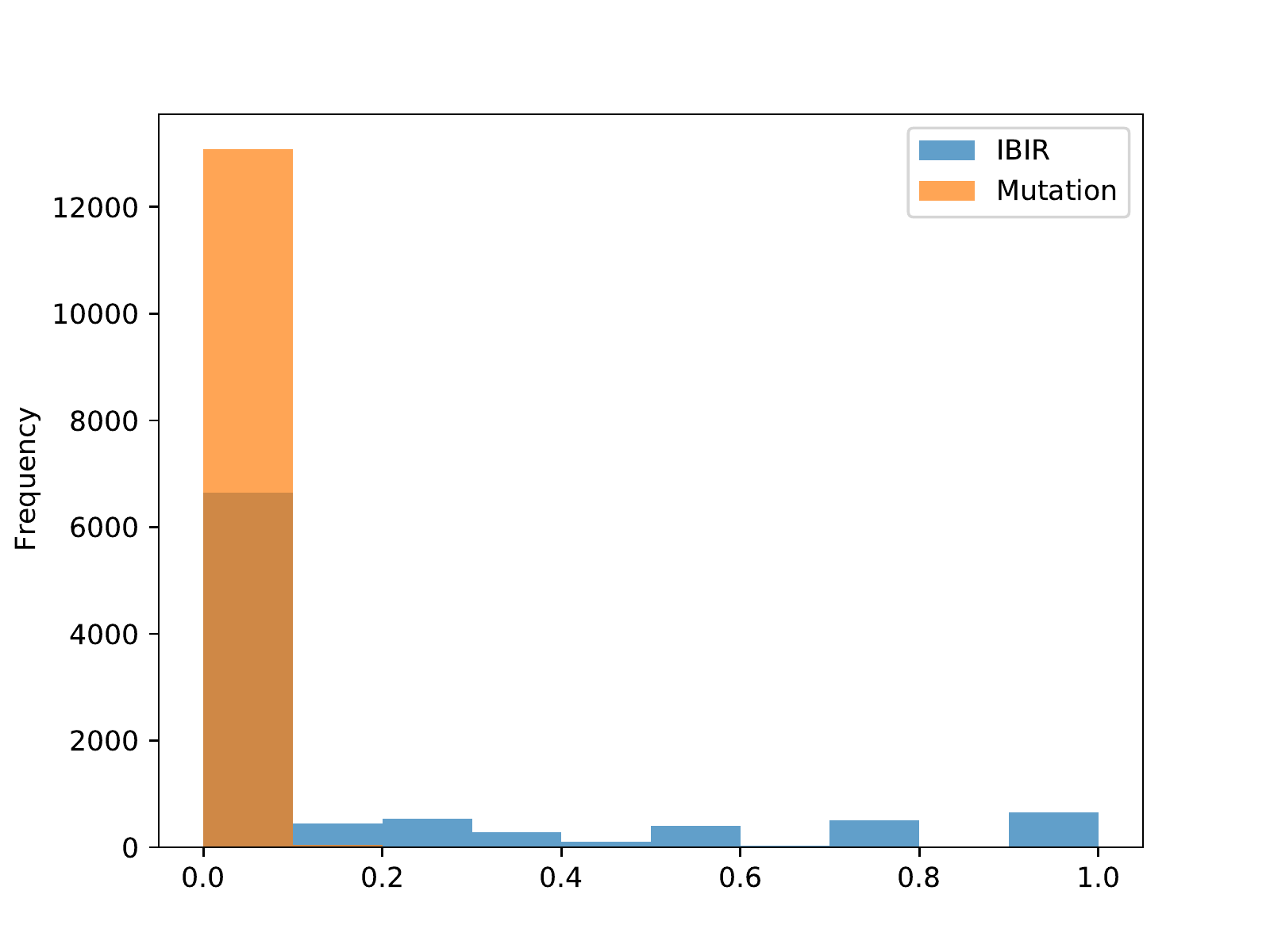}
    \vspace{-0.5em}
    \caption{Distribution of semantic similarities of 100 injected faults per targeted (real) fault. }
    \label{fig:SimilarityDistribution}
\end{figure}

\begin{figure}[t]
\centering 
    \includegraphics[width=0.5\textwidth]{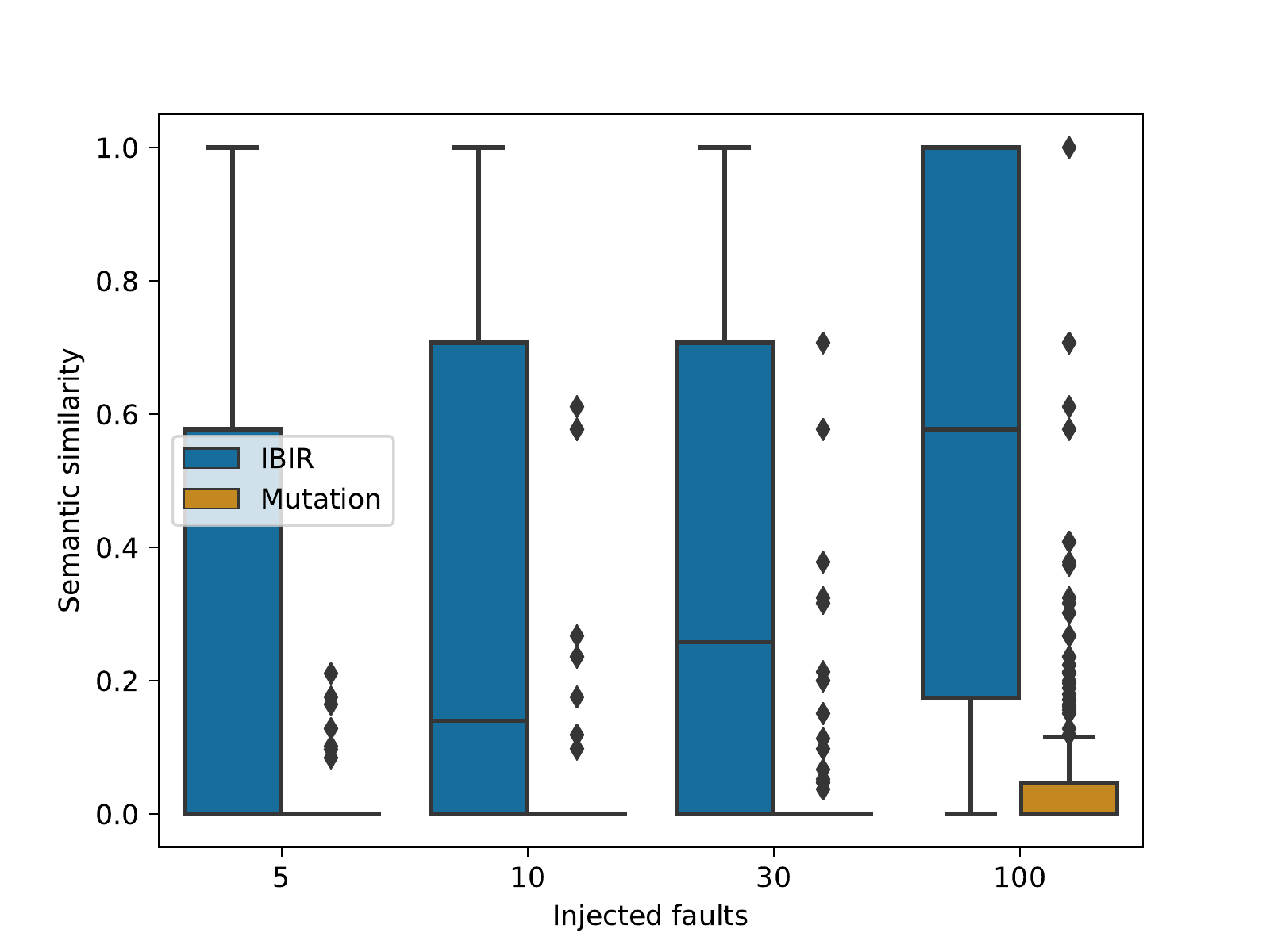}
    \caption{Semantic similarity per targeted (real) fault, top values. \toolname injects faults with higher similarity coefficients than mutation testing.}
    \label{fig:MaxSimilarity}
    \vspace{1.0em}
\end{figure}

To check whether the injected faults imitate well the targeted ones, we measured their behaviour (semantic) similarity w.r.t. the project test suites (please refer to Section \changing{\ref{sec:setup}} for details). Figure \changing{\ref{fig:SimilarityDistribution}} shows the distribution of the similarity coefficient values that were recorded in our study. As can be seen, \toolname injects hundreds of faults that are similar to real ones, whereas mutation (denoted as \changing{Mutation} in Figure \changing{\ref{fig:SimilarityDistribution}}) did not manage to generate any. At the same time, as typically happens in mutation testing \cite{PapadakisSYB18}, a large number of injected faults have low similarity. This is evident in our data, where mutations have 0 similarity.

To investigate whether \toolname successfully injects any fault that is similar (semantically) to the targeted ones, we collected the best similarity coefficients, per targeted fault, when injecting 5, 10, 30 and 100 faults. Figure \changing{\ref{fig:MaxSimilarity}} shows the distribution of these results. For more than half of the targeted faults, \toolname yields a best similarity value higher than 0.5, when injecting 100 faults, indicating that \toolname's faults imitate relatively well the targeted ones. We also observe that in many faults the best similarity values are above 0 by injecting just 10 faults. This is important since it indicates that \toolname successfully identifies relevant locations for fault injection. 

To establish a baseline and better understand the value of \toolname, we need to contrast  \toolname's performance with that of mutation testing when injecting the same number of faults. Mutation testing forms the current SoA of fault injection and thus a related baseline. As can be seen from Figure \changing{\ref{fig:MaxSimilarity}}, the similarity values of mutation testing are significantly lower than those of \toolname.  
In the following subsection we further compare \toolname with mutation testing. 

\subsection{RQ2: \toolname Vs Mutation Testing}

\begin{figure}[t]
\centering 
    \includegraphics[width=0.5\textwidth]{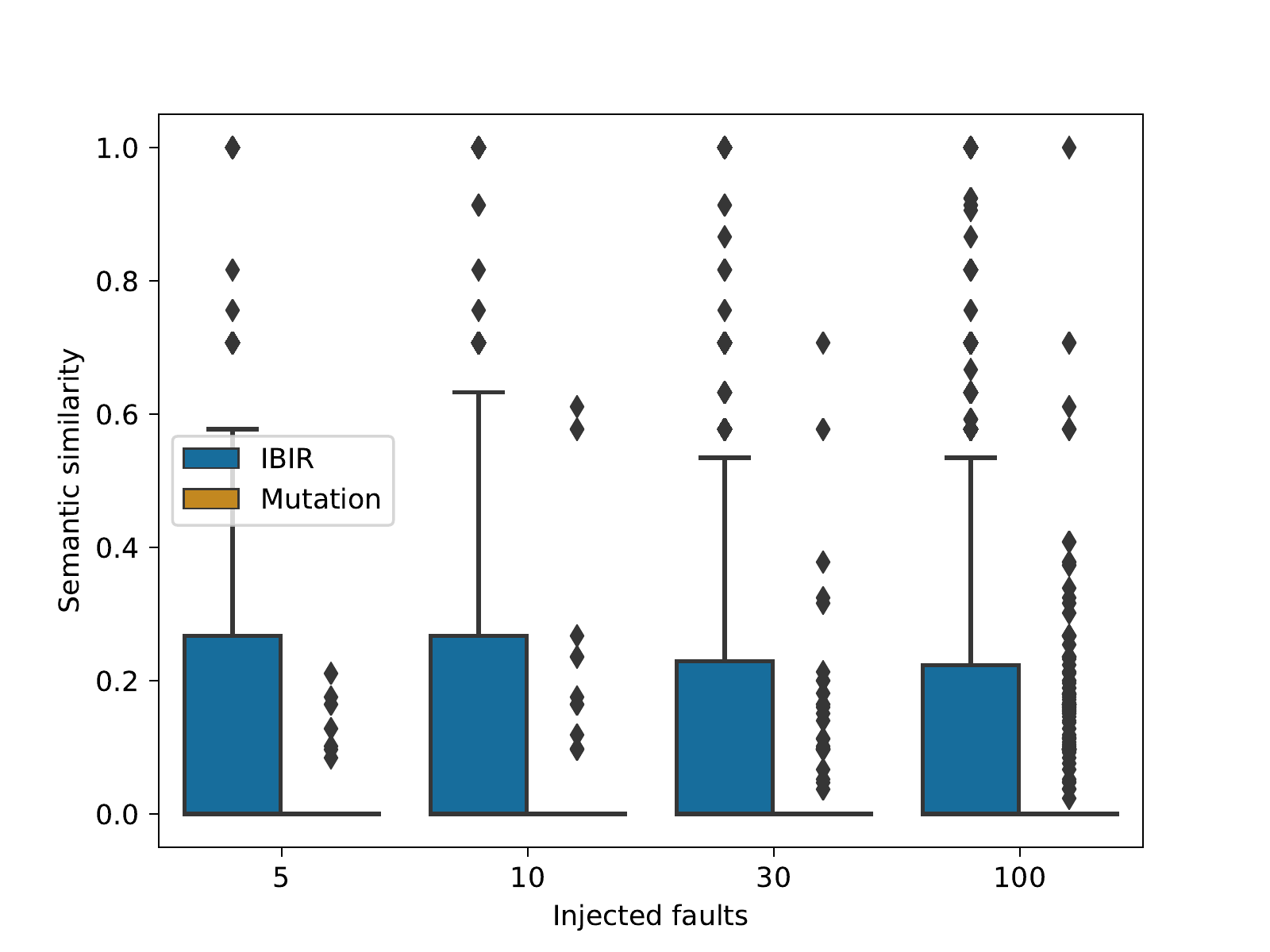}
    \caption{Semantic similarity of all injected faults. \toolname injects faults with higher similarity coefficients than mutation testing.}
    \label{fig:Similarity}
\end{figure}

Figure \changing{\ref{fig:Similarity}} shows the distribution of the semantic similarities, between real and injected faults, when injecting 5, 10, 30 and 100 faults. As can be seen from the boxplots, the trend is that a large portion of faults injected by \toolname imitates the targeted ones, (at least much better than mutation testing). Interestingly, in mutation testing, only outliers have their similarity above 0. In particular, mutation testing injected faults with similarity values higher than 0 in \changing{3, 8, 19, 40} of the targeted faults (when injecting 5, 10, 30, 100 faults), while \toolname injected in \changing{75, 88, 101, 123} of the targeted faults, respectively. 

To validate this finding, we performed a statistical test (Wilcoxon paired test) on the data of both figures \changing{\ref{fig:MaxSimilarity}} and \changing{\ref{fig:Similarity}} to check for significant differences. Our results showed that the differences are significant, indicating the low probability of this effect to be happening by chance. The size of the difference is also big, with \toolname yielding  $\hat{\text{A}}_{12}$ values between \changing{0.73} and \changing{0.84} indicating that \toolname injects faults with higher semantic similarity to real ones in the great majority of the cases. Due to the many cases with 0 similarity values, the average similarity values of \toolname's faults is \changing{0.166}, while for mutation it is \changing{0.002}, indicating the superiority of \toolname.

\subsection{RQ3: \toolname Vs Mutation Testing at particular classes}

\begin{figure}[t]
\centering 
\vspace{-0.5em}
    \includegraphics[width=0.5\textwidth]{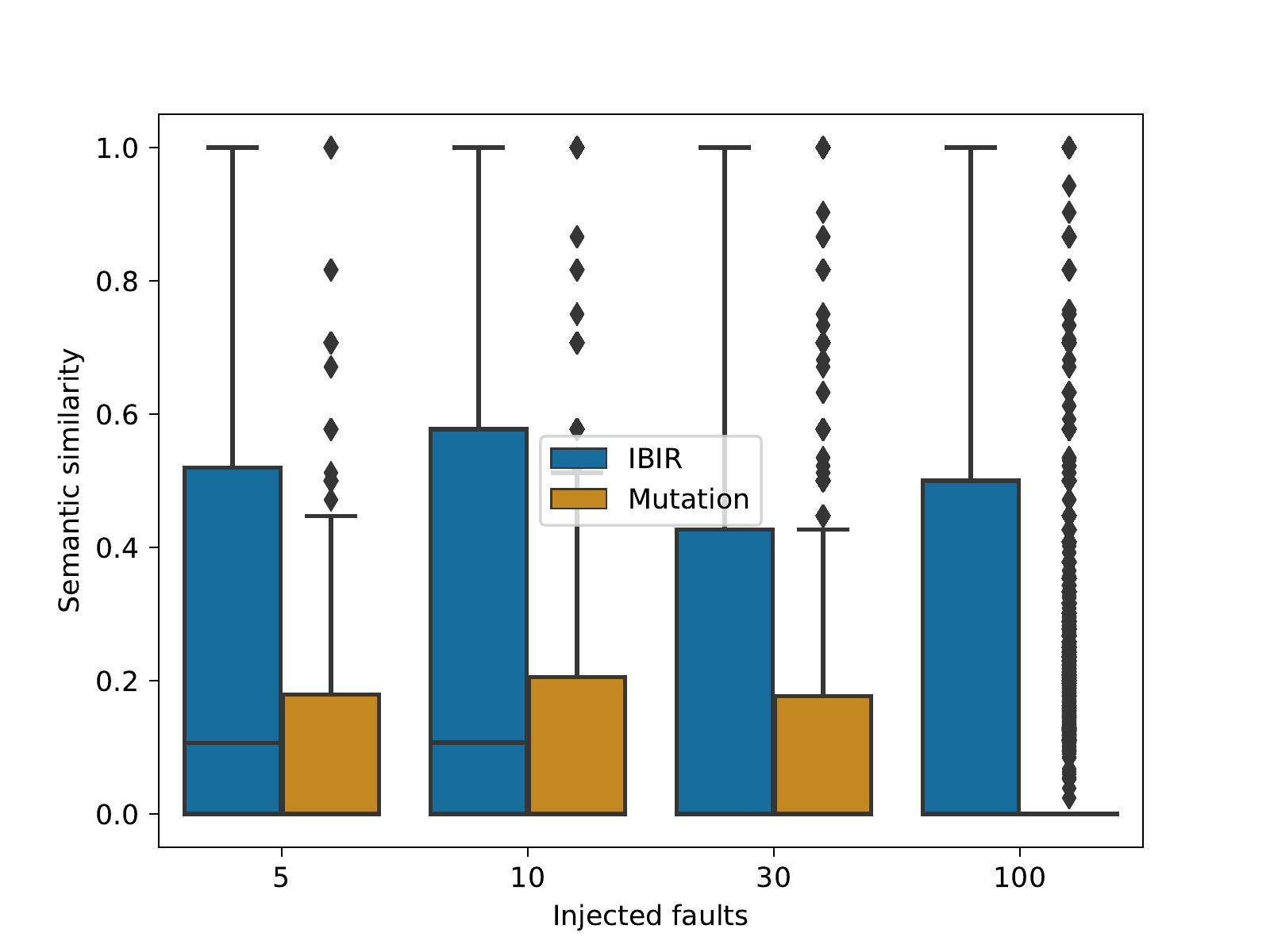}
    \caption{Semantic similarity of injected faults at particular classes. \toolname injects faults with higher similarity coefficients than mutation testing.}
    \label{fig:SimilarityClass}
    \vspace{-0.5em}
\end{figure}

To check the performance of \toolname at the class level of granularity we repeated our analysis by discarding, from our priority lists, every mutant that is not located on the targeted classes, i.e., classes where the targeted faults have been fixed.  Figure \changing{\ref{fig:SimilarityClass}} shows the distribution of the semantic similarities when injecting 5, 10, 30 and 100 faults at a particular class. As expected, mutation testing scores are higher than those presented before, but still mutation testing falls behind. 

To validate this finding, we performed a statistical test and found that the differences are significant. The size of the difference is 0.6, meaning that \toolname score \changing{60\%}  times higher than mutation testing. The average similarity values of the \toolname faults is \changing{0.240}, while for mutation is \changing{0.114}, indicating that \toolname is better.

\subsection{RQ4: Fault Coupling}

\begin{figure}[t]
\centering 
\vspace{-1.0em}
    \includegraphics[width=0.5\textwidth]{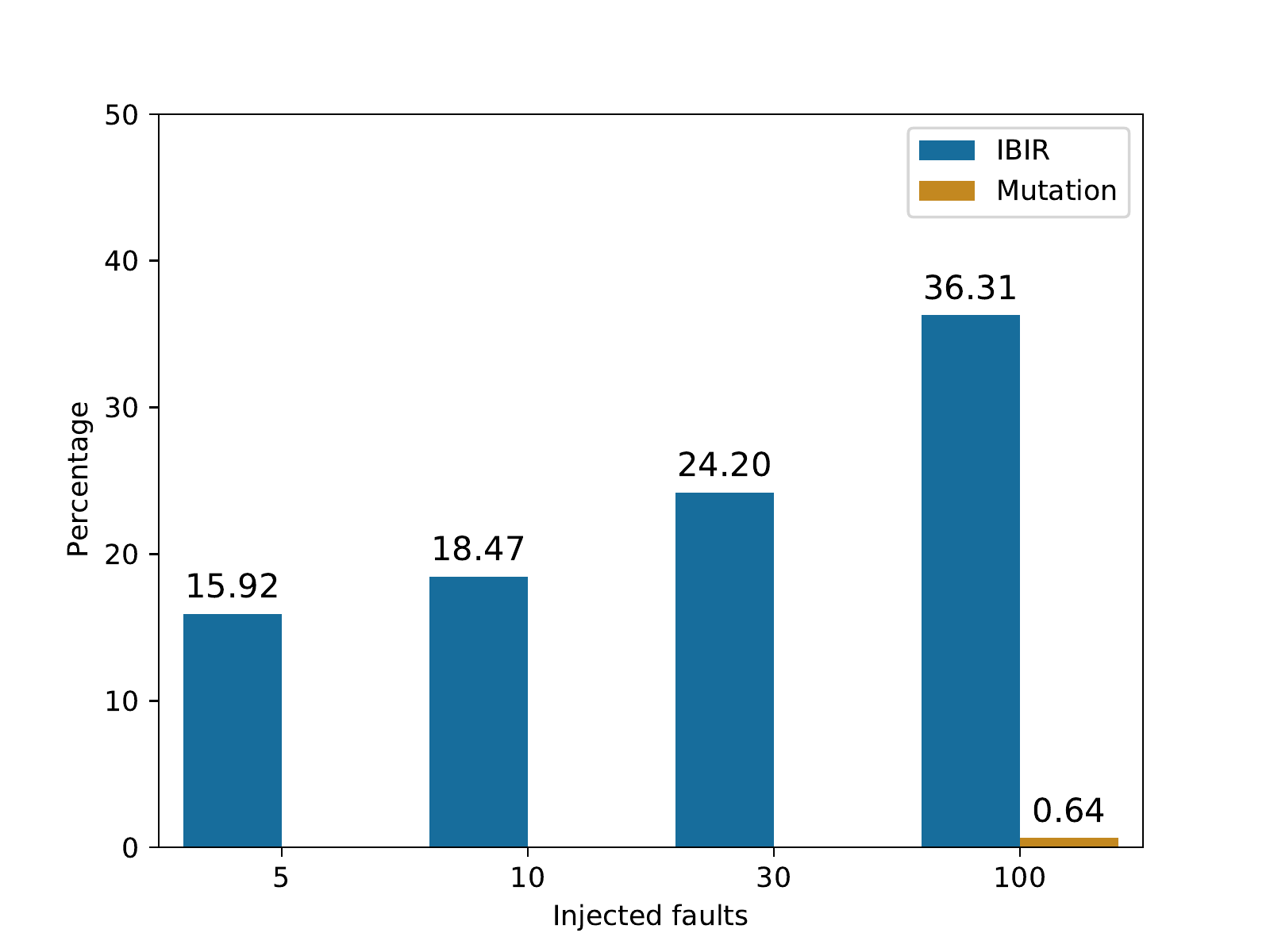}
    \caption{Percentage of injected faults that are coupled to the real ones.}
    \label{fig:FaultCoupling}
    \vspace{-0.5em}
\end{figure}

The coupling between the injected and the real faults forms a fundamental assumption of the fault-based testing approaches \cite{just2014defects4j}. An injected fault is coupled to a real one when a test case that reveals the injected fault also reveals the real fault \cite{just2014defects4j}. This implies that revealing these coupled injected faults results in revealing potential real ones. We therefore, check this property in the faults we inject and contrast it with the baseline mutation testing approach. 

Figure \changing{\ref{fig:FaultCoupling}} shows the percentage of targeted faults where there is at least one injected fault that is coupled to a real one. This is shown for the scenarios where 5, 10, 30 and 100 faults, per target, are injected. As we can see from these data, \toolname injects coupled faults for approximately \changing{16\%}  of the target faults when it aims at injecting 5 faults. This percentage increases to \changing{36\%} when the number of injected faults is increased to 100. 

Perhaps surprisingly, mutation testing did not perform well (it injected coupled faults for \changing{less than 1\%} of the targeted, when injecting 100 faults per target). These results differ from those reported by previous research \cite{JustJIEHF14,PapadakisSYB18}, because a) previous research only injected faults at the faulty classes and not the entire project and b) previous research injected all possible mutant instances and not 100 as we do.

\input{vda-correlation-table} 

\subsection{RQ5: Fault detection estimates}

The results presented so far provide evidence that some of the injected faults imitate well the targeted ones. Though, the question of whether the injections provide representative results of real faults remains, especially since we observe a large number of faults with low similarity value. Therefore, we check the correlations between the failure rates of the sets of injected faults and the real faults when executed with different test suites, (please refer to section \changing{\ref{sec:setup}} for details). 

\begin{figure}[t]
\centering 
\vspace{-1.0em}
    \includegraphics[width=0.5\textwidth]{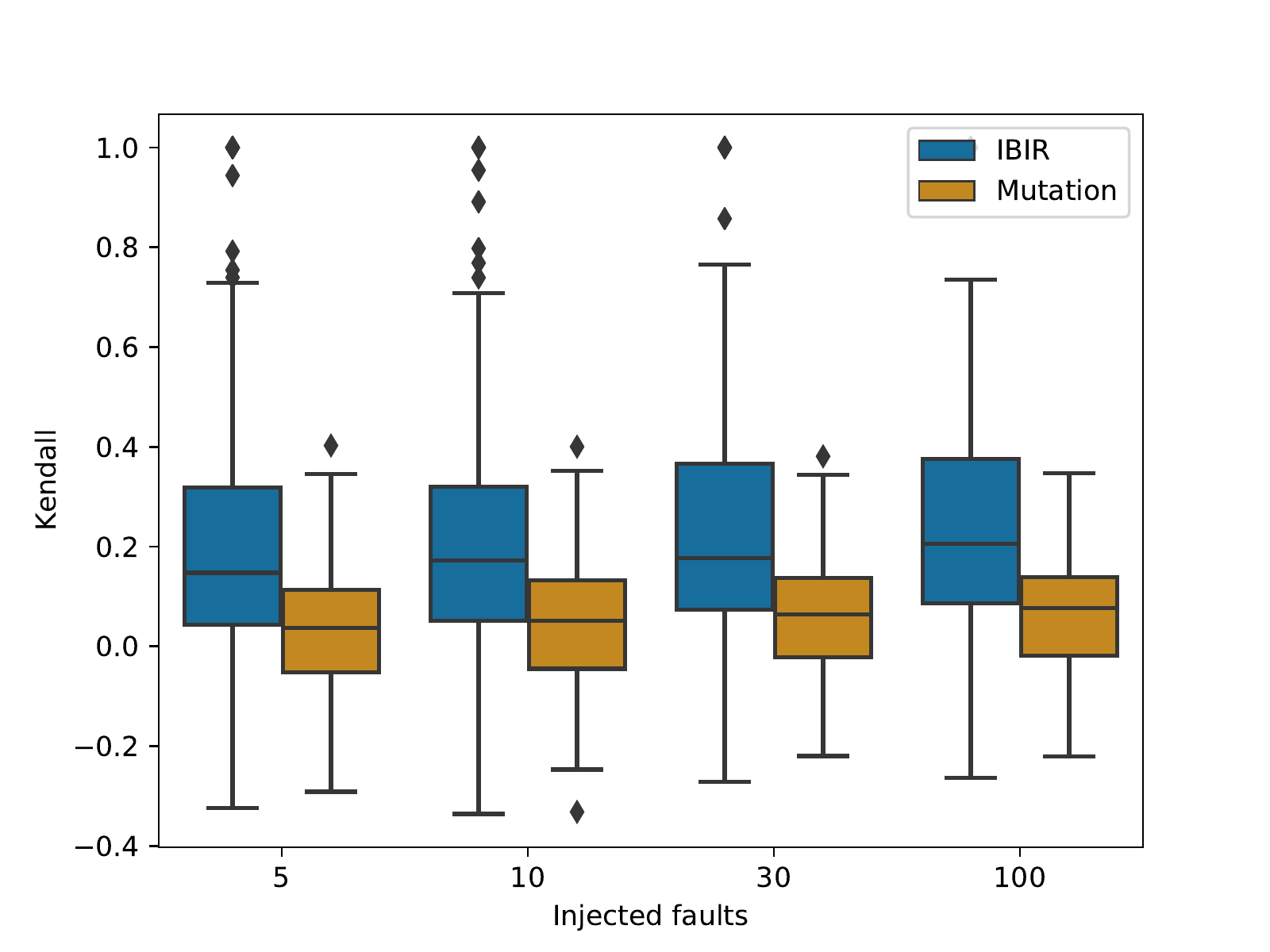}
     \vspace{-2.0em}
    \caption{ Kendall correlation coefficients of test suites (samples from the original project test suite). The two related variables are a) the percentage of injected faults that was detected by the sampled test suites and b) whether the targeted fault was detected or not by the same test suites.}
    \label{fig:Kendall}
    \vspace{-0.5em}
\end{figure}

\begin{figure}[t]
\centering 
\vspace{-1.0em}
    \includegraphics[width=0.5\textwidth]{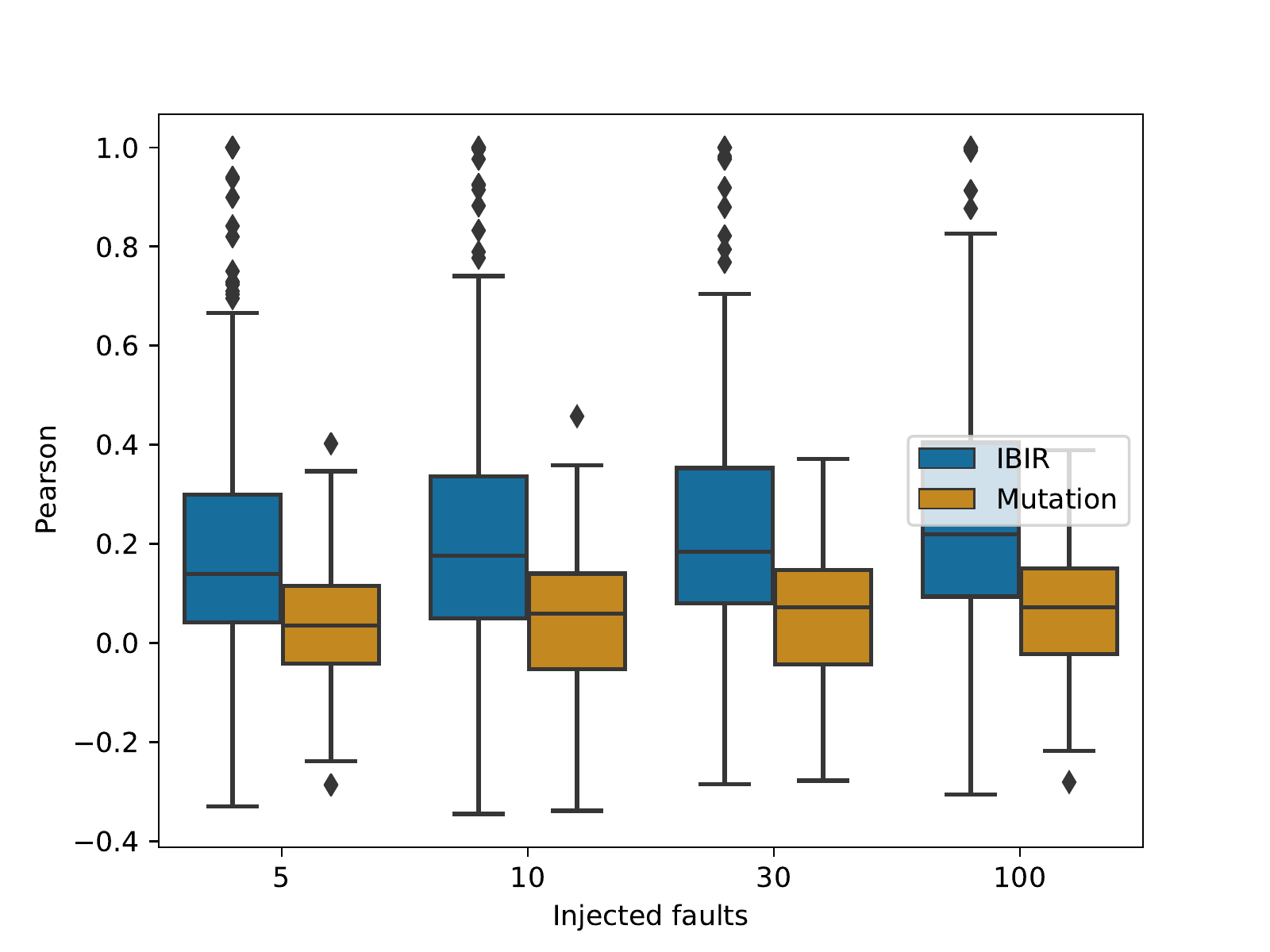}
    \vspace{-2.0em}
    \caption{ Pearson correlation coefficients of test suites (samples from the original project test suite). The two related variables are a) the percentage of injected faults that was detected by the sampled test suites and b) whether the targeted fault was detected or not by the same test suites.}
    \label{fig:Pearson}
    \vspace{-0.5em}
\end{figure}

Figures  \changing{\ref{fig:Kendall}} and  \changing{\ref{fig:Pearson}} show the distribution of the correlation coefficients, when injecting different numbers of faults. Interestingly, the results on both figures show a trend in favour of \toolname. This difference is statistically significant, shown by a Wilcoxon test, with an effect size of approximately \changing{0.72}. Table \ref{tab:VDA-Correlations} records the effect size values, $\hat{\text{A}}_{12}$, for the examined strategies. In essence, these effect sizes mean that \toolname outperforms the mutant injection in 72\% of the cases, suggesting that \toolname could be a much better choice than mutation testing, especially in cases of large test suites with expensive test executions.

To further validate whether \toolname's faults provide good indicators (estimates) of test effectiveness (fault detection) we split our test suites between those that detect the targeted faults and those that do not. We then tested whether detection ratios of the injected faults in the test suite group that detects the real faults are significantly (statistically) higher than those in the group that does not detect it. In case this happens, we can conclude that test suites capable of detecting a higher number of injected faults have similarly higher chances to detect the real ones. This is important when comparing test generation techniques, where the aim is to identify the most effective (at detecting faults) technique. 

\begin{figure}[t]
\centering 
\vspace{-1.0em}
    \includegraphics[width=0.5\textwidth]{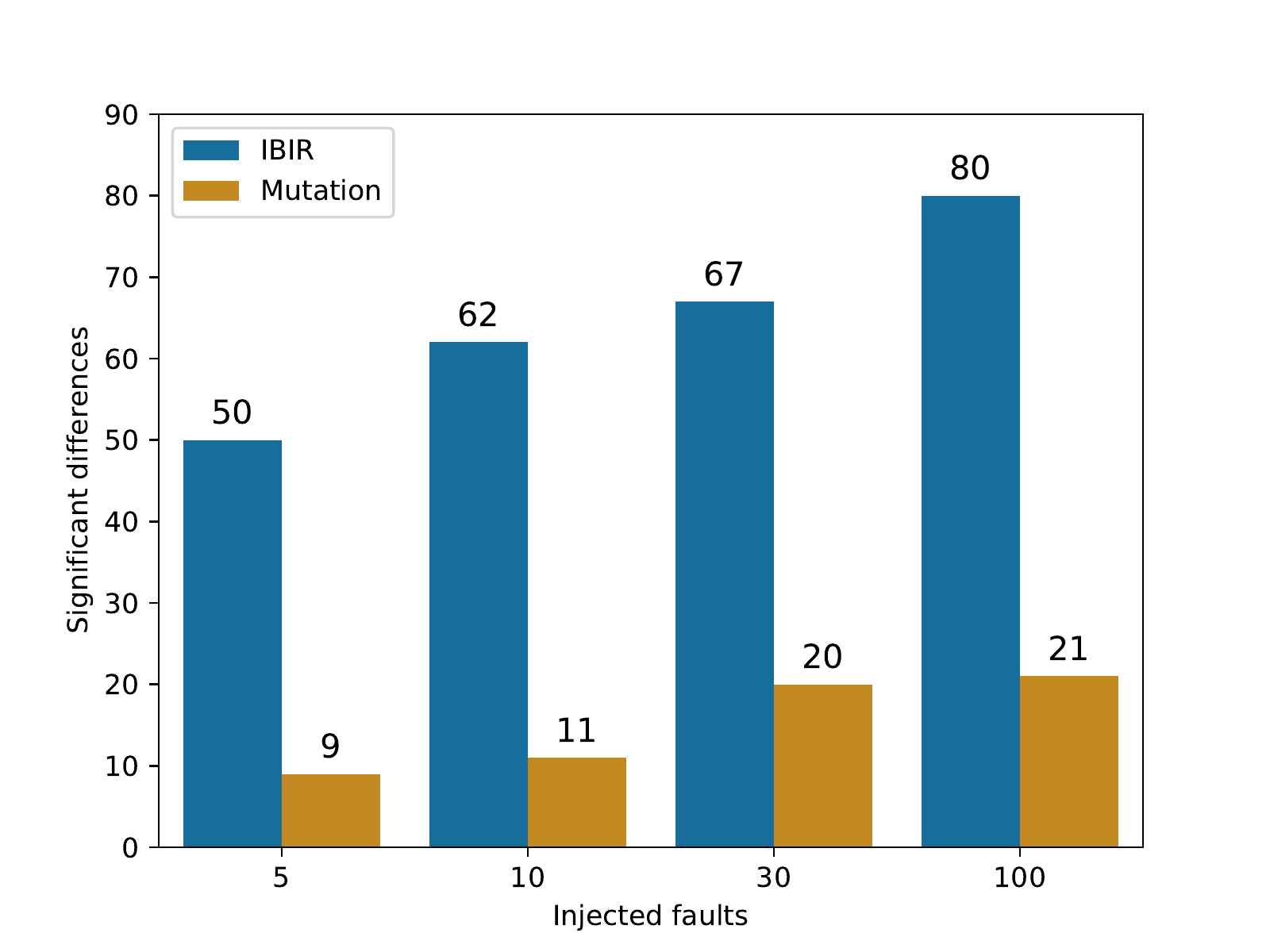}
    \vspace{-2.0em}
    \caption{Number of faults where injected faults provided good indications of fault detection. Particularly, number of cases with test suites detecting the real fault have statistically significant difference, in terms of ratios of injected faults detected, from those that do not detect the real fault.}
    \label{fig:TestEffectivness}
    \vspace{-0.5em}
\end{figure}

\begin{figure}[t]

\centering 
\vspace{-0.5em}
    \includegraphics[width=0.5\textwidth]{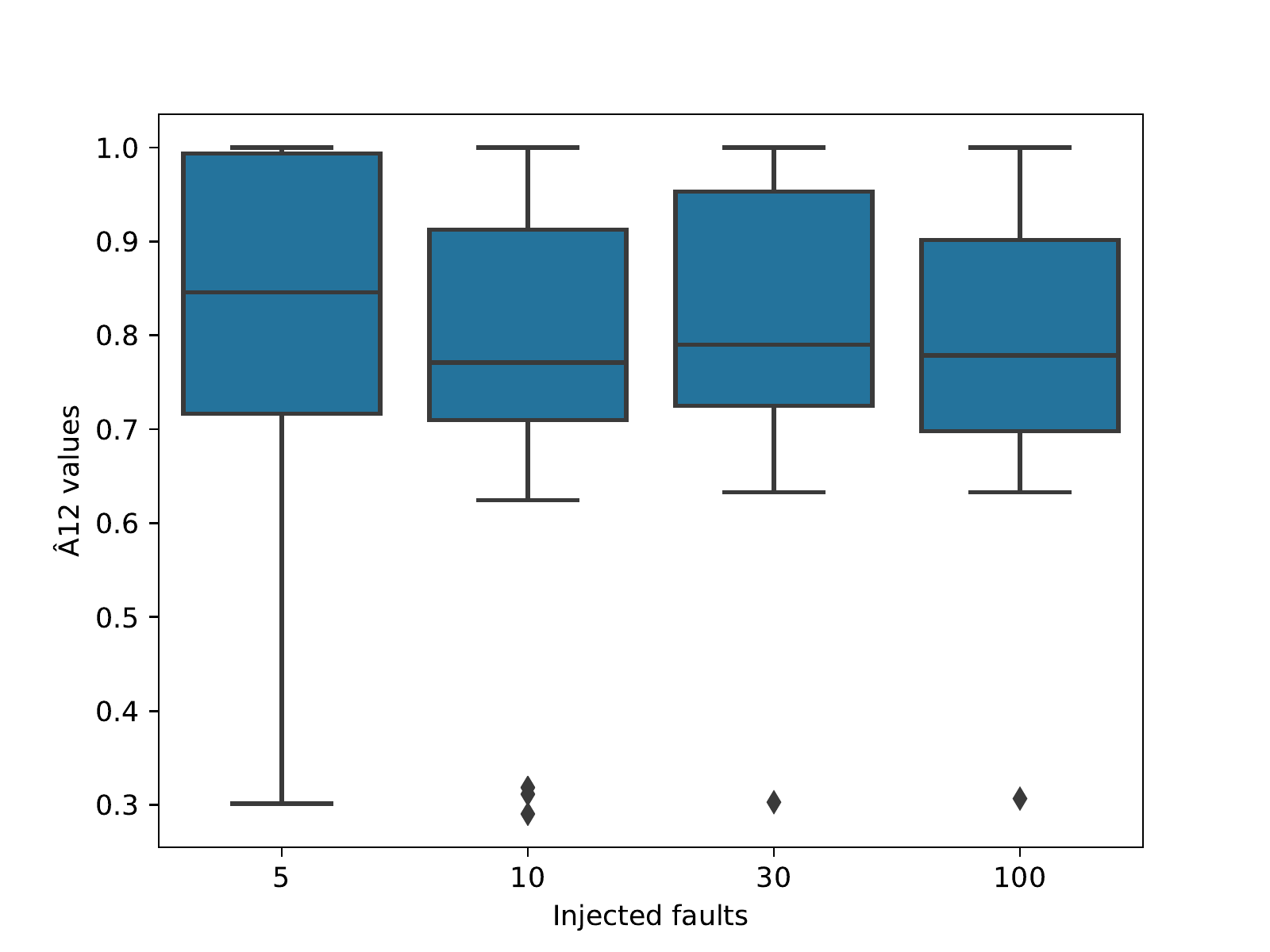}
    \vspace{-2.0em}
\caption{Vargha and Deianey values for \toolname. $\hat{\text{A}}_{12}$ values computed on the detection ratios of injected faults of the test suites that detect and do not detect the (real) faults.}    
    \label{fig:Vargha}
    \vspace{-0.5em}
\end{figure}

Figure \ref{fig:TestEffectivness} records the number of faults where the test suites detecting the (real) targeted fault also detect a statistically higher number of injected faults than those test suites that do not detect it. As can be seen by these results, \toolname has a big difference from mutation, i.e., it distinguishes between passing and failing test suites in \changing{80} faults, while Mutation in \changing{21} faults. Since statistical significance does not imply practical significance, we also measured the Vargha and Delaney $\hat{\text{A}}_{12}$ effect size values on the same data, recorded in Figure \ref{fig:Vargha}. Of course it does not make sense to contrast insignificant cases, so we only performed that on the results where \toolname has statistically significant difference. Interestingly the results demonstrate big differences (in approximately 80\% of the cases) in favour of our approach.

%% file: vda-correlation-table.tex
\begin{table}[t]
\centering
\caption{Vargha and Deianey $\hat{\text{A}}_{12}$ (\toolname VS Mutation) of Kendall and Pearson correlation coefficients. }
\begin{tabular}{@{}lllll@{}}
\toprule
\textbf{Number of injected faults}        & \textbf{5}         & \textbf{10}        & \textbf{30}        & \textbf{100}       \\ \midrule
\textbf{Kendall} & 0.720  & 0.756 & 0.725 & 0.756 \\ \midrule
\textbf{Pearson} & 0.726  & 0.737 & 0.744 & 0.788 \\ \bottomrule
\end{tabular}
\label{tab:VDA-Correlations}
\vspace{-0.5em}
\end{table}

%% file: 7-Threats.tex
\section{Threats to Validity}
\label{sec:Threats}

The question of whether our findings generalise, forms a typical threat to validity of empirical studies. To reduce this threat, we used real-world projects, developer test suites, real faults and their associated bug reports, from an established and independently built benchmark. Still though, we have to acknowledge that these may not be representative of projects from other domains or industrial systems. 

Other threats may also arise from the way we handled the injected faults and mutants that were not killed by any test case. We believe that this validation process is sufficient since the test suites are relatively strong and somehow form the current state of practice, i.e., developers tend to use this particular level of testing. Though, in case the approach is putted into practice things might be different. We also applied our analysis on the fixed program version provided by Defects4J. This was important in order to show that we actually inject the actual targeted faults. Though, our results might not hold on the cases that the code has drastically changed since the time of the bug report. We believe that this threat is not of actual importance as we are concerned with fault injection at interesting program locations, which should be pinpointed by the fault localization technique we use. Still future research should shed some light on how useful these locations and faults are. 

Finally, our evaluation metrics may induce some additional threats. Our comparison basis measurement, i.e., number of injected faults, approximates the execution cost of the techniques and their chances to provide misleading guidance \cite{PapadakisSYB18}, while the fault couplings and semantic similarity metrics approximate the effectiveness of the approaches. These are intuitive metrics, used by previous research \cite{KurtzAODKG16, ChekamPBTS20} and aim at providing a common ground for comparison.   

%% file: 8-RW.tex
\section{Related Work}
\label{sec:RW}


Software fault injection \cite{10.5555/275587} has been widely studied since 1970s. Injected faults have been used for the purpose of testing \cite{PapadakisK00TH19}, debugging \cite{papadakis2015metallaxis, LouGLZZHZ20}, assessing fault tolerance \cite{NatellaCDM13}, risk analysis \cite{ChristmanssonC96, VoasCMMF97} and dependability evaluation \cite{DArlatCCLP93}. 

Despite the many years of research, the majority of previous research is focused on the fault types. In mutation testing research, mutation operators (fault types) are usually designed based on the grammar of the targeted language \cite{0020331, PapadakisK00TH19}, which are then refined through empirical analysis, aiming at reducing the redundancy between the injected faults \cite{OffuttLRUZ96, MarcozziBKPPC18}. The most prominent mutant selection approach is that of Offutt et al. \cite{OffuttLRUZ96}, which proposed a set of 5 mutation operators. This set has been incorporated in most of the modern mutation testing tools \cite{KintisPPVMT18} and is the one that we use in our baseline. 

Recently, Brown et al. \cite{BrownVLR17} aimed at inferring fault patterns from bug fixes. Their results showed that a large number of mutation operators could be inferred. Along the same lines Tufano et al. \cite{TufanoWBPWP19} developed a neural machine translation tool that learns to mutate through bug fixes. 
A key assumptions of these methods are a) the availability of a comprehensive number of clean bug fixing commits, and b) the absence of fault couplings \cite{Offutt92}, which are often not met and can often be reduced to what simple mutations do. For instance, the study of Brown et al. found that with few exceptions, almost all mutation operators designed based on the C language grammar appeared in the inferred operator set. Perhaps more importantly, the studies of Natella et al. \cite{NatellaCDM13} and Chekam et al. \cite{ChekamPBTS20} found that the pair of mutant location and type are what makes mutants powerful and not the type itself. Nevertheless, \toolname goal is complementary to the above studies as it aims at injecting faults that mimic specifically targeted faults, those described in bug reports. This way, one can inject the most important and severe faults experienced. 

Some studies attempt to identify the program locations where to inject faults. Sun et al. \cite{SunXLZ17} suggested injecting faults in diverse places within different program execution paths. Gong et al. \cite{GongZYM17} used graph analysis to inject faults in different and diverse locations of the program spectra. Mirshokraie et al. \cite{Mirshokraie0P15} employed complexity metrics together with actual program executions to inject faults at places with good observability. These strategies, aim at reducing the number of injected faults and not to mimic any real fault as our approach. Moreover, their results should be resembled by the random mutant sampling baseline that we use. 

Random mutant sampling forms a natural cost-reduction method proposed since the early days of mutation testing \cite{DeMilloLS78}. Despite that, most of the mutant selection methods fail to perform better than it. Recently, Kurtz et al. \cite{KurtzAODKG16} and Chekam et al. \cite{ChekamPBTS20} demonstrated that selective mutation and random mutant sampling perform similarly. From this, it should be clear that despite the advances in selective mutation, the simple random sampling is one of the most effective fault injection techniques. This is the reason why we adopt it as a baseline in our experiments.

Natella et al. \cite{NatellaCDM13} used complexity metrics as machine learning features and applied them on a set of examples in order to identify (predict) which injected faults have the potential to emulate well the behaviour of real ones. Chekam et al. \cite{ChekamPBTS20} also used machine learning, with many static mutant-related features to select and rank mutants that are likely fault revealing (have high chance to couple with a fault). These studies assume the availability of a historical faults and do not aim at injecting specific faults as done by \toolname. 

The relationship between injected and real faults has also received some attention \cite{PapadakisK00TH19}. The studies of Papadakis et al. \cite{PapadakisSYB18}, Just et al. \cite{JustJIEHF14}, Andrews et al. \cite{AndrewsBLN06} investigated whether mutant kills and fault detection ratios follow similar trends. The results show the existence of a correlation and, thus, that mutants can be used in controlled experiments as alternatives to real faults. In the context of testing, i.e., using mutants to guide testing, injected faults can help identifying corner cases and reveal existing faults. The studies of Frankl et al. \cite{FranklWH97}, Li et al. \cite{LiPO09} and Chekam et al. \cite{ChekamPTH17} demonstrated that guidance from mutants leads to significantly higher fault revelation than that of other test techniques (test criteria).


%% file: 9-conclusion.tex
\section{Conclusion}
\label{sec:conclusion}

We presented \toolname; a bug-report driven fault injection tool. 
\toolname (1) equips researchers with faults (to inject) targeting the critical functionality of the target systems, (2) mimics real faulty behaviour and (3) makes relevant fault injection. 

\toolname's use case is simple; given a program and some carefully selected bug reports, it injects faults emulating the related bugs, i.e., \toolname generates few faults per target bug report. This allows constructing realistic fault pools to be used for test or fault tolerance assessment.

This means that \toolname's faults can be used as substitutes of real faults, in controlled studies. In a sense, \toolname can bring the missing realism into fault injection and therefore support empirical research and controlled experiments. This is important since a large number of empirical studies rely on artificially-injected faults \cite{PapadakisHHJT16}, the validity of which is always in question.

While the use case of \toolname is in research studies, the use of the tool can have applications in a wide range of software engineering tasks. It can, for instance, be used for asserting that future software releases do not introduce the same (or similar) kind of faults. Such a situation occurs in large software projects~\cite{palix2011faults}, where \toolname could help by checking for some of the most severe faults experienced.

Another potential application of \toolname is fault tolerance assessment, by injecting faults similar to previously experienced ones and  analysing the system responses and overall dependability.

Finally, testers could use \toolname for testing all system areas that could lead to similar symptoms than the ones observed and resolved. This will bring significant benefits when testing software clones \cite{MondalRSRKS11} and similar functionality implementations. We hope that we will address these points in the near future.